\documentclass[prd,preprint,superscriptaddress]{revtex4}
\usepackage{amssymb}
\usepackage{revsymb}
\usepackage{graphicx}

\begin{document}

\title{Relic gravitational waves and present accelerated expansion}
\author{Germ\'{a}n Izquierdo \thanks{%
E-mail address: german.izquierdo@uab.es}}
\affiliation{Departamento de F\'{\i}sica, Universidad Aut\'{o}noma de Barcelona, 08193
Bellaterra (Barcelona), Spain}
\author{Diego Pav\'{o}n\thanks{%
E-mail address: diego.pavon@uab.es}}
\affiliation{Departamento de F\'{\i}sica, Universidad Aut\'{o}noma de Barcelona, 08193
Bellaterra (Barcelona), Spain}

\begin{abstract}
We calculate the current power spectrum of the gravitational waves created
at the big bang (and later amplified by the different transitions during the
Universe expansion) taking into account the present stage of accelerated
expansion. Likewise, we determine the power spectrum in a hypothetical
second dust era that would follow the present one if at some future time the
dark energy, that supposedly drives the current accelerated expansion,
evolved in such a way that it became dynamically equivalent to cold dark
matter. The calculated power spectrums as well as the evolution of the
density parameter of the waves may serve to discriminate between different 
phases of expansion and may help ascertain the nature of dark energy.
\end{abstract}

\maketitle

\section{Introduction}

Relic gravitational waves (RGWs, for short) are believed to have their
origin in a variety of mechanisms at the earliest instants of the big bang.
Of particular importance is parametric amplification but there are others 
proposed mechanisms that might also prove very efficients in this respect \cite%
{ks,alessandra,bruce}. Under certain conditions, when the Universe made a
transition from a stage of expansion dominated by a given energy source to
the next, e.g., inflation--radiation era, radiation--matter era, these
primodial waves experienced amplification. This has been widely studied by
Grishchuk \cite{leonid}, Allen \cite{Allen} and Maia \cite{Maia93}. Assuming
that a mini black hole dominated phase existed right after reheating and
prior to radiation dominance two other transitions may have taken place
(namely, inflation--mini black holes era and mini black holes--radiation
era) with the corresponding amplifications bearing a profound impact on the
final power spectrum \cite{german}. Since the cosmic medium is nearly
transparent to gravitational waves \cite{transparent} their detection by
LISA \cite{lisa} or some other suitable antenna is expected to yield
invaluable information on the earliest epochs of the Universe expansion and
may allow us to reconstruct the whole history of the scale factor.

Nowadays the observational data of the luminosity of supernovae type Ia
strongly suggests that the expansion of the Universe is accelerated at
present \cite{snia}. In Einstein gravity this is commonly associated either
to a cosmological constant (vacuum energy) or to a sort of energy, the
so--called dark energy, that violates the strong energy condition and
clusters only at the largest accesible scales \cite{iap}. In such a case the
present state of the Universe would be dominated by dark energy and since it
redshifts more slowly with expansion than dust, the contribution of the
latter is bound to become negligible. Some models, however, propose dark
energy potentials such that the current acceleration phase would be just
transitory and sooner or later the expansion would revert to the
Einstein--de Sitter law, $a(t) \propto t^{2/3}$, thereby slowing down
(second dust era) \cite{Alam}.

The aim, of this paper is to show how the power spectrum as well as the
dimensionless density parameter of the gravitational waves created at the
big bang, and later enhanced by parametric amplifications may help present
(and future) observers to ascertain whether the expansion phase they are
living in is accelarated or not and if accelerated, which law is obeying the scale
factor. The latter would facilitate enormously to discrimate the nature of
dark energy between a large variety of proposed models (cosmological
constant, quintessence fields, interacting quintessence, tachyon fields,
Chaplygin gas, etc) \cite{iap}. To this end we calculate the power spectrum
and energy density of the RGWs when the transitions to the dark energy
era and second dust era are considered. Obviously, the latter power spectrum
lies at the future and depending on the model under consideration it may
take very long for the Universe to enter the second dust era. In general,
the scale factor of the FRW metric will be of the form 
$a(\eta )\propto\eta^{n}$, where $\eta$ is the conformal 
time and $n$ a constant parameter (but different in each 
expansion phase). Accordingly, we are implicitly assuming 
that in each era there is a constant relation between 
pressure and energy density.

Section II briefly describes the different eras of expansion and obtains the
Bogoliubov coefficients of each transition. Section III calculates the power
spectrum of the RGWs at the present era of accelerated expansion as well as
the power spectrum at the second dust era. Section IV determines the
evolution of the energy density of gravitational waves within the Hubble
radius from the present era onwards. Finally, Section V summarizes our
findings. The notation and conventions of Ref. \cite{german} are assumed
throughout.

\section{Transitions and coefficients of Bogoliubov}

We consider a spatially flat FRW scenario initially De Sitter, then
dominated  by radiation, followed by a dust dominated era, an accelerated
expansion era domiated by dark energy, and finally a second dust era. The
scale factor in terms of the conformal time reads%
\begin{equation}
a(\eta )=\left\{ 
\begin{array}{c}
-\frac{1}{H_{1}\eta }\qquad (-\infty <\eta <\eta _{1}<0),\qquad \text{%
inflation era} \\ 
\frac{1}{H_{1}\eta _{1}^{2}}(\eta -2\eta _{1})\qquad (\eta _{1}<\eta <\eta
_{2}),\qquad \text{radiation era} \\ 
\frac{1}{4H_{1}\eta _{1}^{2}}\frac{(\eta +\eta _{2}-4\eta _{1})^{2}}{\eta
_{2}-2\eta _{1}}{\qquad }(\eta _{2}<\eta <\eta _{3}),\qquad \text{first dust
era} \\ 
\left( \frac{l}{2}\right) ^{-l}\frac{(\eta _{3}+\eta _{2}-4\eta _{1})^{2-l}}{%
4H_{1}\eta _{1}^{2}(\eta _{2}-2\eta _{1})}\left( \eta _{l}\right) ^{l}{%
\qquad }(\eta _{3}<\eta <\eta _{4}),\qquad \text{dark energy era} \\ 
\frac{a_{4}}{4}\left( a_{4}H_{4}\right) ^{2}\left( \eta -\eta _{4}+\frac{2}{%
a_{4}H_{4}}\right) ^{2}{\qquad }(\eta _{4}<\eta ),\qquad \text{second dust
era}%
\end{array}%
\right.  \label{sclfac}
\end{equation}%
where $l\leq -1$, $\eta _{l}=\eta +\frac{l}{2}\left[ \left( -2/l+1\right)
\eta _{3}+\eta_{2}-4\eta _{1}\right]$, the subindexes $1,2,3,4$ correspond
to sudden transitions from inflation to radiation era, from radiation to
first dust era, from first dust era to dark energy era and from the latter
to the second dust era, respectively, $H_{i}$ is the Hubble factor at the
instant $\eta = \eta_{i}$. The present time $\eta _{0}$ lies in the range $%
\left[ \eta _{3},\eta _{4}\right] $, it is to say in the dark energy
dominated era. 

\begin{figure}[tbp]
\includegraphics*[angle=-90,scale=0.7]{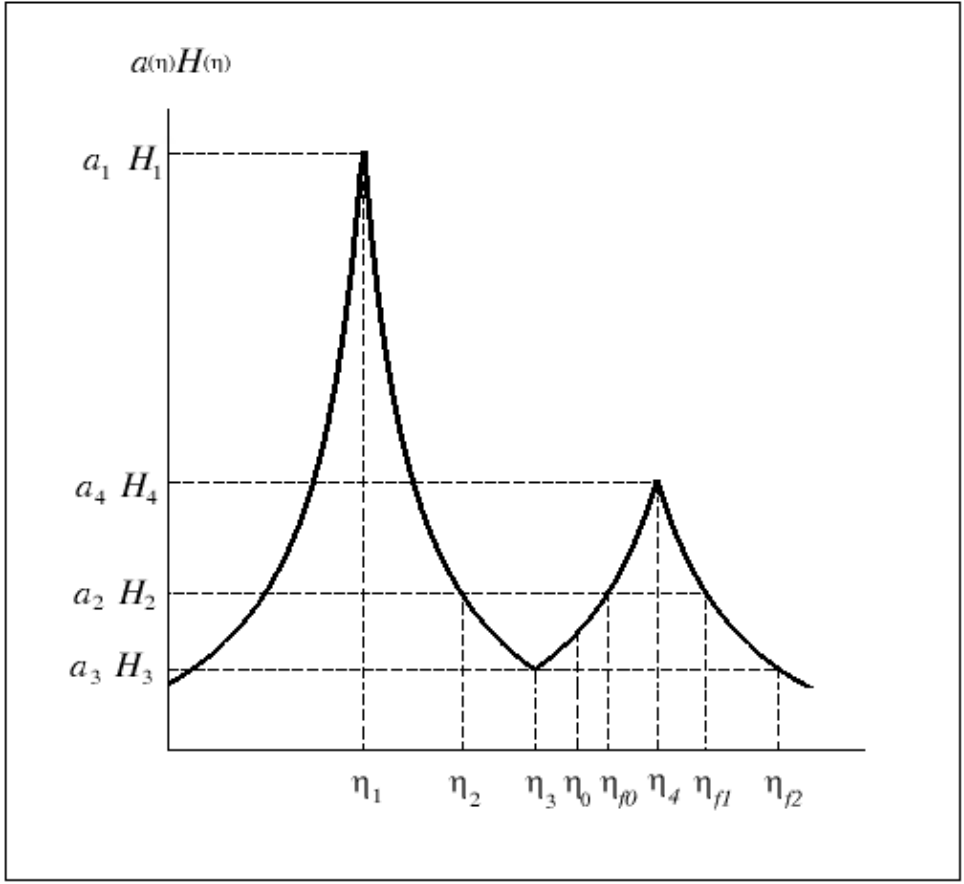}
\caption{{}Evolution of $a(\protect\eta )H(\protect\eta )$ in a universe
with scale factor given by Eq. (\protect\ref{sclfac}). We have assumed that $%
a_{4}H_{4}>a_{2}H_{2}$. The quantity $\protect\eta _{f0}$ is defined as the
instant in the dark energy era in which $a(\protect\eta _{f0})H(\protect\eta %
_{f0})=a_{2}H_{2}$, $\protect\eta _{f1}$ as the instant of the second dust
era such that $a(\protect\eta _{f1})H(\protect\eta _{f1})=a_{2}H_{2},$ and $%
\protect\eta _{f2}$ as the instant of the second dust era in which $a(%
\protect\eta _{f0})H(\protect\eta _{f0})=a_{3}H_{3}$. }
\label{aH}
\end{figure}

Figure \ref{aH} schematically shows the evolution of $a(\eta )H(\eta )$.
During the inflationary and dark energy eras $a(\eta )H(\eta )$ increases
with $\eta $, and decreases in the other eras. As a consequence, $a_{4}H_{4}$
results higher than $a_{3}H_{3}$. Choosing $l$, $\eta _{3}$ and $\eta _{4}$
in such a way that

\begin{equation}
\left( \frac{a_{4}}{a_{3}}\right) ^{-1/l}\left( \frac{a_{2}}{a_{3}}\right)
^{1/2}>1,  \label{cond}
\end{equation}%
we have that $a_{4}H_{4}$ is also higher than $a_{2}H_{2}$. We assume that $%
a_{4}H_{4}$ is lower than $a_{1}H_{1}$ throughout. 

The gravitational waves equation may be interpreted as the massless Klein-Gordon 
equation \cite{Allen, Parker}. Its solution can be written as%
\begin{equation}
h_{ij}(\eta ,\mathbf{x})=\int \left( A_{(k)}h_{ij}^{(k)}(\mathbf{k}%
,x)+A_{(k)}^{\dag }h_{ij}^{(k)}(\mathbf{k},x)\right) d^{3}k,  \label{fieldh}
\end{equation}%
\[
h_{ij}^{(k)}(\mathbf{k},x)=\frac{1}{\sqrt{\pi }}e_{ij}(\mathbf{k})\frac{\mu
_{(k)}(\eta )}{a(\eta )}e^{i\mathbf{k\cdot x}}\mathbf{,} 
\]%
\begin{equation}
\mu _{(k)}^{\prime \prime }(\eta )+\left( k^{2}-\frac{a^{\prime \prime
}(\eta )}{a(\eta )}\right) \mu _{(k)}(\eta )=0,  \label{eqmu}
\end{equation}%
where $A_{(k)}$, $A_{(k)}^{\dag }$ are the annihilation and creation
operators, respectively, $e_{ij}(\mathbf{k})$ contains the two possible
polarizations of the wave and the modes $\mu _{(k)}(\eta )$ satisfy the
additional condition%
\begin{equation}
\mu _{(k)}\mu _{(k)}^{\ast \prime }-\mu _{(k)}^{\ast }\mu _{(k)}^{\prime }=i,
\label{condcuan}
\end{equation}%
which comes from the commutation relations of the operators $A_{(k)}$, $%
A_{(k)}^{\dag }$ and the definition of the field $h_{ij}(\eta ,\mathbf{x})$ 
\cite{Parker, Birrell}.

To study the amplification of gravitational waves at the transition eras we must 
ascertain which of the solutions to (\ref{eqmu}) correspond to real particles. An
approach to that problem is known as the adiabatic vacuum approximation \cite%
{Birrell}. Basically it assumes that in the limit $k\rightarrow \infty $ the
spacetime becomes asymptotically Mikowskian. In this limit the
creation-destruction operators of each family of modes will correspond
exactly to those associated to real particles.

Following Allen \cite{Allen} and Maia \cite{Maia93}, we shall evaluate the
number of RGWs created from the initial vacuum state in an expanding
universe. Assuming that the scale factor in the different eras is given by
Eq. (\ref{sclfac}), the initial state is the vacuum associated with the
modes of the inflationary stage $\mu _{I}(\eta )$, which are a solution to
the equation (\ref{eqmu}) compatible with condition (\ref{condcuan}). Taking
into account the shape of the scale factor at this era the modes are 
\begin{equation}
\mu _{I}=(\sqrt{\pi }/2)e^{i\psi _{I}}k^{-1/2}x^{1/2}H_{-3/2}^{(2)}(x),
\label{muinfcuant}
\end{equation}%
where $x=k\eta $, $\psi _{I}$ an arbitrary constant phase and $%
H_{-3/2}^{(2)}(x)$ the Hankel function of order $-3/2$ \cite{german}.
Likewise, the proper modes in the radiation era are%
\begin{equation}
\mu _{R}=(\sqrt{\pi }/2)e^{i\psi
_{R}}k^{-1/2}x_{R}^{1/2}H_{1/2}^{(2)}(x_{R}),  \label{murad}
\end{equation}%
where $x_{R}=k(\eta -2\eta _{1})$ and $\psi _{R}$ a constant phase.

The two families of modes are related by the Bogoliubov transformation%
\begin{equation}
\mu _{I}(\eta )=\alpha _{1}\mu _{R}(\eta )+\beta _{1}\mu _{R}^{\ast }(\eta ).
\label{bogtr}
\end{equation}%
From the continuity of $\mu (\eta )$ at the transition time $\eta _{1}$, it
follows that 
\begin{equation}
\alpha _{1}=-1+\frac{i}{k\eta _{1}}+\frac{1}{2(k\eta _{1})^{2}},\qquad \beta
_{1}=\frac{1}{2(k\eta _{1})^{2}},  \label{coef1}
\end{equation}%
where we have neglected an irrelevant phase. The modes with frequency $%
f=2\pi k/a(\eta _{1})$ at $\eta_{1}$ larger than the characteristic time
scale of the transition are exponentially suppressed. This time scale is
usually identified with the inverse of the Hubble factor at the transition, $%
H_{1}^{-1}$ in this case. Therefore, the Bogoliubov coefficients are given
by $\alpha_{1}=1$ and $\beta _{1}=0$ for RGWs with $k>2\pi a_{1}H_{1}$, and
by Eq. (\ref{coef1}) when $k<2\pi a_{1}H_{1}$.

In the first dust era ($\eta _{2}<\eta <\eta _{3}$) the solution to Eq. (\ref%
{eqmu}) for the modes is found to be 
\begin{equation}
\mu _{D}=(\sqrt{\pi }/2)e^{i\psi
_{D}}k^{-1/2}x_{D}^{1/2}H_{3/2}^{(2)}(x_{D}),  \label{mudustcuan}
\end{equation}%
where $x_{D}=k\left( \eta +\eta _{2}-4\eta _{1}\right) $ and it is related
to the radiation modes by

\begin{equation}
\mu _{R}(\eta )=\alpha _{2}\mu _{D}(\eta )+\beta _{2}\mu _{D}^{\ast }(\eta ).
\label{bogtrr_d}
\end{equation}%
Consequently, one obtains%
\begin{equation}
\alpha _{2}=-i\left( 1+\frac{i}{2k\eta _{R2}}-\frac{1}{8\left( k\eta
_{R2}\right) ^{2}}\right) ,\qquad \beta _{2}=\frac{i}{8\left( k\eta
_{R2}\right) ^{2}},  \label{coef2}
\end{equation}%
for $k<2\pi a(\eta _{2})H_{2}$ and $\alpha _{2}=1$, $\beta _{2}=0$ for $%
k>2\pi a(\eta _{2})H_{2}$, where $H_{2}$ is the Hubble factor evaluated at $%
\eta _{2}$ and $\eta _{R2}=\eta _{2}-2\eta _{1}$.

Finally, in the dark energy era the solution to equation (\ref{eqmu}) reads 
\[
\mu _{l}=(\sqrt{\pi }/2)e^{i\psi
_{l}}k^{-1/2}x_{l}^{1/2}H_{l-1/2}^{(2)}(x_{l}), 
\]%
where $x_{l}=k\eta _{l}$. The Bogoliubov transformation%
\begin{equation}
\mu _{D}(\eta )=\alpha _{3}^{l}\mu _{l}(\eta )+\beta _{3}^{l}\mu _{l}^{\ast}
(\eta )  \label{bogtrr_l}
\end{equation}%
relates the dust and dark energy modes.

Using the well-known relation \cite{Abr}%
\begin{equation}
H_{l-1/2}^{(2)}(x)\simeq e^{-ix}\sqrt{\frac{2}{\pi x}}\frac{(-2l)!}{(-l)!}%
(-2)^{l}x^{l}(1+ix),  \label{aproxhankl}
\end{equation}%
valid when $x\ll 1$ and $l<-1$, it follows that%
\begin{eqnarray*}
\alpha _{3} &=&\left( 6(-1)^{-l}l^{l}\frac{(-2l)!}{(-l)!}\right) \eta
_{m3}^{l-2}k_{{}}^{l-2}+\mathcal{O}(k_{{}}^{l-1}), \\
\beta _{3} &=&-\left( 6(-1)^{-l}l^{l}\frac{(-2l)!}{(-l)!}\right) \eta
_{m3}^{l-2}k_{{}}^{l-2}+\mathcal{O}(k_{{}}^{l-1}),
\end{eqnarray*}%
for $k<2\pi a(\eta _{3})H_{3}$ and $\alpha _{3}=1$, $\beta _{3}=0$ for $%
k>2\pi a(\eta _{3})H_{3}$, where $H_{3}$ is the Hubble factor evaluated at $%
\eta _{3}$ and $\eta _{m3}=\eta _{3}+\eta _{2}-4\eta _{1}$. The transition
between the first dust era to the dark energy era has a time span of the
order $H_{3}^{-1}$. This time span is much shorter than the period of the
waves we are considering and therefore it may be assumed instantaneous in
the calculation of the coefficients. As a consequence, the time span from
this transition till today $\tau =t_{0}-t_{3}$ must be larger than $%
H_{3}^{-1}$, otherwise the transition first dust era-dark energy era would
be too close to the present time for our formalism to apply. This condition
places a bound on the value of the redshift $\frac{a_{0}}{a_{3}}$. When $l=-1$
the condition is $\frac{a_{0}}{a_{3}}>2.72$, $\frac{a_{0}}{a_{3}}>2.25$ when 
$l=-2$, $\frac{a_{0}}{a_{3}}>2.15$ when $l=-3$, and $\frac{a_{0}}{a_{3}}>2$
when $l\rightarrow -\infty $. This bound is compatible with the accepted
values for $\frac{a_{0}}{a_{3}}\in \lbrack 1.5,11]$ (see e.g., \cite{luca}).
Henceforward we will consider $\frac{a_{0}}{a_{3}}$ larger than $2.72$ and
no larger than $11$.

The solution to equation (\ref{eqmu}) in the second dust era (i.e., the one
following the dark energy era) is 
\[
\mu _{SD}=(\sqrt{\pi }/2)e^{i\psi
_{SD}}k^{-1/2}x_{SD}^{1/2}H_{3/2}^{(2)}(x_{SD}), 
\]%
where $x_{SD}=k\left( \eta -\eta _{4}+\frac{2}{a_{4}H_{4}}\right) $.

The Bogoliubov coefficients relating the modes of the dark energy era with
the modes of the second dust era are%
\[
\mu _{l}(\eta )=\alpha _{4}^{l}\mu _{SD}(\eta )+\beta _{4}^{l}\mu
_{SD}^{\ast }(\eta ). 
\]

The continuity of $\mu $ at $\eta _{4}$ implies%
\begin{eqnarray*}
\alpha _{4}^{l} &=&\left( -1\right) ^{l}\frac{3}{8}2^{l}\frac{(-2l)!}{(-l)!}%
x_{l4}^{l-2}\left[ l^{2}+il(l-2)x_{l4}^{{}}+ \mathcal{O}(x_{l4}^{2})\right] ,
\\
\beta _{4}^{l} &=&\left( -1\right) ^{l}\frac{3}{8}2^{l}\frac{(-2l)!}{(-l)!}%
x_{l4}^{l-2}\left[ l^{2}+il(l+2)x_{l4}^{{}}+ \mathcal{O}(x_{l4}^{2})\right] ,
\end{eqnarray*}%
for $k<2\pi a_{4}H_{4}$, and $\alpha _{4}^{l}=1$, $\beta _{4}^{l}=0$ for $%
k>2\pi a_{4}H_{4}$, where $H_{4}$ is the Hubble factor evaluated at the
transition time $\eta _{4}$.

\section{Power spectrums}

In order to relate the modes of the inflationary era to the modes of the
first dust era, we make use of the total Bogoliubov coefficients $\alpha
_{Tr}$ and $\beta _{Tr}$. The number of RGWs created from the initial vacuum
is $\left\langle N_{\omega }\right\rangle =\left\vert \beta _{Tr}\right\vert
^{2}$. Assuming that each RGW has an energy $2\hbar \omega (\eta )$
(henceforward we shall use ordinary units), the energy density of RGWs with
frequencies in the range $\left[ \omega (\eta ),\omega (\eta )+d\omega (\eta
)\right] $ can be written as 
\begin{equation}
d\rho _{g}(\eta )=2\hbar \omega (\eta )\left[ \frac{\omega ^{2}(\eta )}{2\pi
^{2}c^{3}}d\omega (\eta )\right] \left\langle N_{\omega }\right\rangle
=P(\omega (\eta ))d\omega (\eta ),  \label{densquan}
\end{equation}%
where $P(\omega (\eta ))=\left( \hbar \omega^{3}(\eta )/\pi ^{2}
c^{3}\right) \left\langle N_{\omega }\right\rangle $ denotes the power
spectrum. As the energy density $\rho _{g}$ is a locally defined quantity,
it loses its meaning for metric perturbations with wave length $\lambda
=2\pi c/\omega (\eta )$ larger than the Hubble radius $c H^{-1}(\eta )$.

\subsection{Current Power spectrum}

To evaluate the present power spectrum one must bear in mind that $%
a_{0}H_{0}>a_{3}H_{3}$ which implies the perturbations created at the
transition dust era-dark energy era have a wave length larger than the
Hubble radius at present \cite{Chiba}. One must also consider the
possibility that $a_{0}H_{0}>a_{2}H_{2}$, it is to say

\[
\left( \frac{a_{0}}{a_{3}}\right) ^{-1/l}\left( \frac{a_{2}}{a_{3}}\right)
^{1/2}>1. 
\]%
When $l=-1$ and assuming $\frac{a_{0}}{a_{2}}\sim 10^{4}$ \cite{Peebles}
this condition implies $\frac{a_{0}}{a_{3}}>21.5$, $\frac{a_{0}}{a_{3}}>100$
when $l=-2,$ and $\frac{a_{0}}{a_{3}}>251.2$ when $l=-3$. The values for $%
\frac{a_{0}}{a_{3}}$ considered by us are larger than $2.72$ and no lower than $%
11$, consequently we can safely assume $a_{0}H_{0}<a_{2}H_{2}$.

For $k>2\pi a_{1}H_{1}$, we find that $\alpha _{Tr2}=1$, $\beta _{Tr2}=0$;
in the range $2\pi a_{1}H_{1}>k>2\pi a_{2}H_{2}$, the coefficients are $%
\alpha _{Tr2}=\alpha _{1}$ and $\beta _{Tr2}=\beta _{1}$, and finally for $%
k<2\pi a(\eta _{2})H_{2}$ we obtain \cite{Allen, Maia93}

\begin{equation}
\beta _{Tr2}\simeq -\frac{1}{8k^{3}\eta _{1}^{2}\eta _{R2}}.  \label{coeft2}
\end{equation}

Thus, the number of RGWs at the present time $\eta _{0}$ created from the
initial vacuum state is $\left\langle N_{\omega }\right\rangle =\left\vert
\beta _{Tr2}\right\vert ^{2}\sim \omega ^{-6}(\eta _{0})$ for $\omega (\eta
_{0})<2\pi (a_{2}/a_{0})H_{2}$, $\omega ^{-4}(\eta _{0})$ for $2\pi
(a_{1}/a_{0})H_{1}>\omega (\eta _{0})>2\pi (a_{2}/a_{0})H_{2}$, and zero for 
$\omega (\eta _{0})>2\pi (a_{1}/a_{0})H_{1}$, where we have used the present
value of the frequency, $\omega (\eta _{0})=k/a_{0}$.

In sumary, the current power spectrum of RGWs in this scenario is 
\begin{equation}
P(\omega )\simeq \left\{ 
\begin{array}{l}
0\text{\qquad }\left( \omega (\eta _{0})>2\pi (a_{1}/a_{0})H_{1}\right) , \\ 
\\ 
\frac{\hbar }{4\pi ^{2}c^{3}}\left( \frac{a_{1}}{a_{0}}\right)
^{4}H_{1}^{4}\omega ^{-1}\text{\qquad }\left( 2\pi (a_{2}/a_{0})H_{2}<\omega
(\eta _{0})<2\pi (a_{1}/a_{0})H_{1}\right) , \\ 
\\ 
\frac{\hbar }{16\pi ^{2}c^{3}}\left( \frac{a_{0}}{a_{2}}\right) ^{2}\left( 
\frac{a_{1}}{a_{0}}\right) ^{8}H_{1}^{6}\omega ^{-3}\text{\qquad }\left(
2\pi H_{0}<\omega (\eta _{0})<2\pi (a_{2}/a_{0})H_{2}\right) .%
\end{array}%
\right.  \label{espectr}
\end{equation}

While this power spectrum is not at variance with the power spectrum of the
conventional three-stage scenario (De Sitter inflation, radiation and dust era) 
\cite{Allen, Maia93}, it evolves differently. The power spectrum in the dark
energy scenario at $\eta =\eta _{3}$ formally coincides with Eq. (\ref%
{espectr}) but with $a_{3}$ substituted by $a_{0}$ throughout, and from then
up to now waves with $2\pi a_{3}H_{3}<k<$ $2\pi a_{0}H_{0}$ cease to
contribute to the spectrum as soon as their wave length exceeds the Hubble
radius. By contrast, in the three-stage scenario waves are continuously
being added to the spectrum. As we shall see in Section IV, this implies
that the evolution of the energy density of the gravitational waves in the
three-stage scenario differs from the scenario in which the Universe
expansion is dominated by dark energy.

\subsection{Power spectrum in the second dust era}

In an attempt to evade the particle horizon problem posed by an everlasting
accelerated expansion to string/M type theories \cite{string/M}, models in
which the present era is not eternal but it is followed by a decelerated
phase as though dominated by cold dark matter (second dust era) were
proposed \cite{Alam}.

Here we evaluate the power spectrum at some future time $\eta $ larger than $%
\eta _{f2}$ for which the waves created at the transition dust era-dark
energy era ($\eta =\eta _{3}$) are considered in the spectrum by the first
time (see Figure \ref{aH}). Let $\alpha _{Tr4}$ and $\beta _{Tr4}$ be the
Bogoliubov coefficients relating the modes of the inflationary era to the
modes of the second dust era. Because of condition (\ref{cond}) we must
consider two possibilities with two different power spectrums.

$(i)$ If condition (\ref{cond}) is not fulfilled, then $%
a_{1}H_{1}>a_{2}H_{2}>a_{4}H_{4}>a_{3}H_{3}$ and the power spectrum can be
obtained from the following total coefficients. In the range $k>2\pi
a_{1}H_{1}$ the total coefficients are $\alpha _{Tr4}=1$ and $\beta _{Tr4}=0$%
. For $2\pi a_{1}H_{1}>k>2\pi a_{2}H_{2}$, the coefficients are $\alpha
_{Tr4}=\alpha _{1}$ and $\beta _{Tr4}=\beta _{1}$, where $\alpha _{1}$ and $%
\beta _{1}$ are defined in Eq. (\ref{coef1}). For $2\pi a_{2}H_{2}>k>2\pi
a_{4}H_{4}$, $\alpha _{Tr4}=\alpha _{Tr2}$ and $\beta _{Tr4}=\beta _{Tr2}$
where $\alpha _{Tr2}$ and $\beta _{Tr2}$ are defined in Eq. (\ref{coeft2}).

When $2\pi a_{4}H_{4}>k>2\pi a_{3}H_{3}$, except for $\alpha _{3}$ and $%
\beta _{3}$, all the cofficients obtained in the previous section must be
considered when evaluating $\alpha _{Tr4}$ and $\beta _{Tr4}$, therefore%
\begin{eqnarray}
\alpha _{Tr4}^{l} &=&\alpha _{4}^{l}\alpha _{Tr2}^{{}}+\beta _{4}^{l\ast
}\beta _{Tr2}^{{}}\simeq \left[ \frac{3}{32}\left( -1\right)
^{-l-1}l^{2}2^{l}\frac{(-2l)!}{(-l)!}\right] \frac{2}{\eta _{1}^{2}\eta
_{R2}\eta _{l4}^{-l+2}}k^{l-5},\qquad  \label{coeft4a} \\
\beta _{Tr4}^{l} &=&\beta _{4}^{l}\alpha _{Tr2}^{{}}+\alpha _{4}^{l\ast
}\beta _{Tr2}^{{}}\simeq \left[ \frac{3}{32}\left( -1\right)
^{-l-1}l^{2}2^{l}\frac{(-2l)!}{(-l)!}\right] \frac{2}{\eta _{1}^{2}\eta
_{R2}\eta _{l4}^{-l+2}}k^{l-5}.  \nonumber
\end{eqnarray}%
Finally, for $2\pi a_{3}H_{3}>k>2\pi H(\eta )$ we get%
\begin{equation}
\beta _{Tr4}^{l}\simeq -\left[ \frac{9}{16}l^{l+2}2^{l}\left( \frac{(-2l)!}{%
(-l)!}\right) ^{2}\right] \frac{2l}{\eta _{1}^{2}\eta _{R2}\eta
_{m3}^{-l}\eta _{l4}^{-l+2}}k^{2l-5}.  \label{coeft4b}
\end{equation}%
Accordingly, the power spectrum reads{\small 
\[
P(\omega )\simeq \left\{ 
\begin{array}{l}
0\text{\qquad }\left( \omega (\eta )>2\pi (a_{1}/a(\eta ))H_{1}\right) , \\ 
\\ 
\frac{\hbar }{4\pi ^{2}c^{3}}\left( \frac{a_{1}}{a(\eta )}\right)
^{4}H_{1}^{4}\omega ^{-1}\text{\qquad }\left( 2\pi (a_{2}/a(\eta
))H_{2}<\omega (\eta )<2\pi (a_{1}/a(\eta ))H_{1}\right) , \\ 
\\ 
\frac{\hbar }{16\pi ^{2}c^{3}}\left( \frac{a_{0}}{a_{2}}\right) ^{2}\left( 
\frac{a_{1}}{a_{0}}\right) ^{8}\left( \frac{a_{0}}{a(\eta )}\right)
^{6}H_{1}^{6}\omega ^{-3}\text{\qquad }\left( 2\pi (a_{4}/a(\eta
))H_{4}<\omega (\eta )<2\pi (a_{2}/a(\eta ))H_{2}\right) \\ 
\\ 
\frac{9\hbar }{\pi ^{2}c^{3}}l^{2l}2^{2l-8}\left( \frac{(-2l)!}{(-l)!}%
\right) ^{2}\left( \frac{a_{1}}{a_{0}}\right) ^{16-4l}\left( \frac{a_{2}}{%
a_{0}}\right) ^{l-4}\left( \frac{a_{3}}{a_{0}}\right) ^{l-4+4/l}\left( \frac{%
a_{4}}{a_{0}}\right) ^{-8+2l-4/l} \\ 
\text{\qquad }\times \left( \frac{a(\eta )}{a_{4}}\right)
^{2l-10}H_{1}^{10-2l}\omega ^{2l-7}\text{\qquad }\left( 2\pi (a_{3}/a(\eta
))H_{3}<\omega (\eta )<2\pi (a_{4}/a(\eta ))H_{4}\right) , \\ 
\\ 
\frac{27\hbar }{\pi ^{2}c^{3}}l^{4l}2^{4l-6}\left( \frac{(-2l)!}{(-l)!}%
\right) ^{4}\left( \frac{a_{1}}{a_{0}}\right) ^{16-8l}\left( \frac{a_{2}}{%
a_{0}}\right) ^{2l-4}\left( \frac{a_{3}}{a_{0}}\right) ^{2l-4+4/l}\left( 
\frac{a_{4}}{a_{0}}\right) ^{4l-8-4/l} \\ 
\text{\qquad }\times \left( \frac{a(\eta )}{a_{4}}\right)
^{4l-10}H_{1}^{10-4l}\omega ^{4l-7}\text{\qquad }\left( 2\pi H(\eta )<\omega
(\eta )<2\pi (a_{3}/a(\eta ))H_{3}\right) .%
\end{array}%
\right. 
\]%
}$(ii)$ If condition (\ref{cond}) is fulfilled, then $%
a_{1}H_{1}>a_{4}H_{4}>a_{2}H_{2}>a_{3}H_{3}$. As in the previous case, in
the range $k>2\pi a_{1}H_{1}$ the total coefficients are $\alpha _{Tr4}=1$
and $\beta _{Tr4}=0$. For $2\pi a_{1}H_{1}>k>2\pi a_{4}H_{4}$, the
coefficients are $\alpha _{Tr4}=\alpha _{1}$ and $\beta _{Tr4}=\beta _{1}$.
In the range $2\pi a_{4}H_{4}>k>2\pi a_{2}H_{2}$, we obtain%
\begin{eqnarray*}
\alpha _{Tr4}^{l} &=&\alpha _{4}^{l}\alpha _{1}+\beta _{4}^{l\ast }\beta
_{1}\simeq i\frac{3}{8}\left( -1\right) ^{-l}l^{2}2^{l}\frac{(-2l)!}{(-l)!}%
\frac{1}{\eta _{1}^{2}\eta _{l4}^{-l+1}}k^{l-3},\qquad \\
\beta _{Tr4}^{l} &=&\beta _{4}^{l}\alpha _{1}+\alpha _{4}^{l\ast }\beta
_{1}\simeq i\frac{3}{8}\left( -1\right) ^{-l}l^{2}2^{l}\frac{(-2l)!}{(-l)!}%
\frac{1}{\eta _{1}^{2}\eta _{l4}^{-l+1}}k^{l-3}.
\end{eqnarray*}%
Again, for $2\pi a_{2}H_{2}>k>2\pi a_{3}H_{3}$, the total coefficients are
given by Eq. (\ref{coeft4a}) while for $2\pi a_{3}H_{3}>k>2\pi H(\eta )$
they obey Eq. (\ref{coeft4b}). The power spectrum in this case is{\small 
\begin{equation}
P(\omega )\simeq \left\{ 
\begin{array}{l}
0\text{\qquad }\left( \omega (\eta )>2\pi (a_{1}/a(\eta ))H_{1}\right) , \\ 
\\ 
\frac{\hbar }{4\pi ^{2}c^{3}}\left( \frac{a_{1}}{a(\eta )}\right)
^{4}H_{1}^{4}\omega ^{-1}\text{\qquad }\left( 2\pi (a_{4}/a(\eta
))H_{4}<\omega (\eta )<2\pi (a_{1}/a(\eta ))H_{1}\right) , \\ 
\\ 
\frac{9\hbar }{\pi ^{2}c^{3}}l^{2l+2}2^{2l-6}\left( \frac{(-2l)!}{(-l)!}%
\right) ^{2}\left( \frac{a_{1}}{a_{0}}\right) ^{8-4l}\left( \frac{a_{2}}{%
a_{0}}\right) ^{l-1}\left( \frac{a_{3}}{a_{0}}\right) ^{l-3+2/l}\left( \frac{%
a_{4}}{a_{0}}\right) ^{2l-4-4/l} \\ 
\text{\qquad } \times \left( \frac{a(\eta )}{a_{4}}\right)
^{2l-6}H_{1}^{6-2l}\omega ^{2l-3}\text{\qquad }\left( 2\pi (a_{2}/a(\eta
))H_{2}<\omega (\eta )<2\pi (a_{4}/a(\eta ))H_{4}\right) , \\ 
\\ 
\frac{9\hbar }{\pi ^{2}c^{3}}l^{2l}2^{2l-8}\left( \frac{(-2l)!}{(-l)!}%
\right) ^{2}\left( \frac{a_{1}}{a_{0}}\right) ^{16-4l}\left( \frac{a_{2}}{%
a_{0}}\right) ^{l-4}\left( \frac{a_{3}}{a_{0}}\right) ^{l-4+4/l}\left( \frac{%
a_{4}}{a_{0}}\right) ^{2l-8-4/l} \\ 
\text{\qquad } \times \left( \frac{a(\eta )}{a_{4}}\right)
^{2l-10}H_{1}^{10-2l}\omega ^{2l-7}\text{\qquad }\left( 2\pi (a_{3}/a(\eta
))H_{3}<\omega (\eta )<2\pi (a_{2}/a(\eta ))H_{2}\right) , \\ 
\\ 
\frac{27\hbar }{\pi ^{2}c^{3}}l^{4l}2^{4l-6}\left( \frac{(-2l)!}{(-l)!}%
\right) ^{4}\left( \frac{a_{1}}{a_{0}}\right) ^{16-8l}\left( \frac{a_{2}}{%
a_{0}}\right) ^{2l-4}\left( \frac{a_{3}}{a_{0}}\right) ^{2l-4+4/l}\left( 
\frac{a_{4}}{a_{0}}\right) ^{4l-8-4/l} \\ 
\text{\qquad }\times \left( \frac{a(\eta )}{a_{4}}\right)
^{4l-10}H_{1}^{10-4l}\omega ^{4l-7}\text{\qquad }\left( 2\pi H(\eta )<\omega
(\eta )<2\pi (a_{3}/a(\eta ))H_{3}\right) .%
\end{array}%
\right.  \label{Ps}
\end{equation}%
} The power spectrum governed by Eq.(\ref{Ps}) is plotted in Fig. \ref%
{spec} for different choices of the free parameters $l$, $\frac{a_{0}}{a_{3}}
$, $\frac{a_{4}}{a_{0}}$ and $\frac{a(\eta)}{a_{0}}$ as well as the power
spectrum assuming the three-stage model, i.e., non-accelerated phase and no
second dust era. 

The shape of the power spectrum given by (\ref{Ps}) in the
range $2\pi H(\eta )<\omega (\eta )<2\pi (a_{3}/a(\eta ))H_{3}$ is the same
in both cases as in this range all the coefficients are present in the
evaluation of the total coefficients. It is interesting to see how markedly
this spectrum differs from the one arising in the three-stage model (dot-dashed
line) at low wave lengths.

\begin{figure}[tbp]
\includegraphics*[scale=0.7]{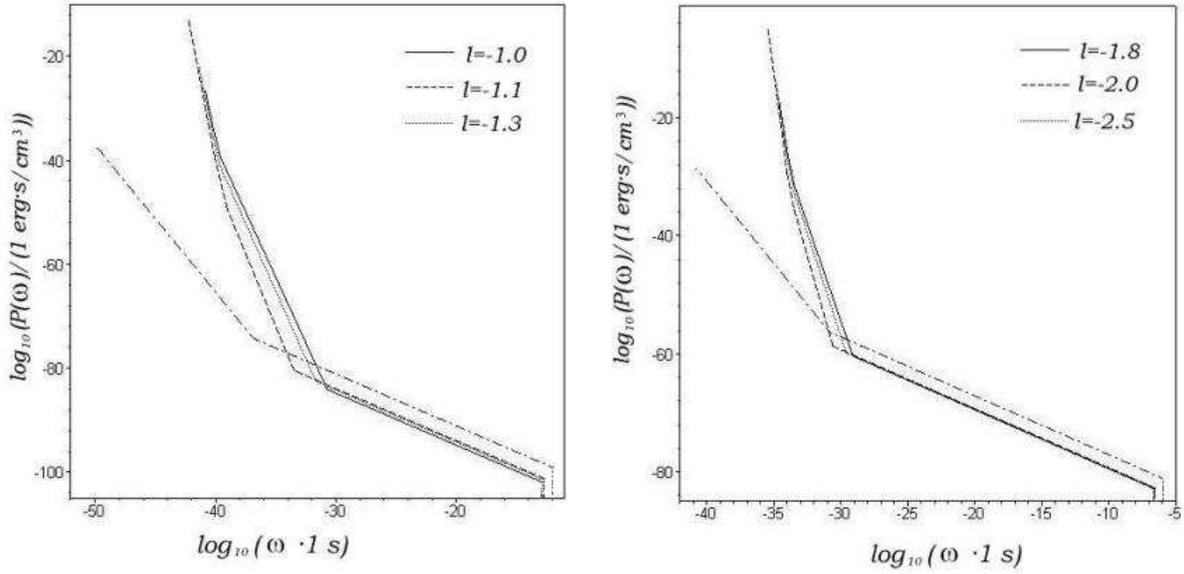}
\caption{Power spectrum given by Eq. (\protect\ref{Ps}) for different values
of $l$. The left panel depicts the power spectrum at the instant $\protect%
\eta $ for which $\frac{a\left(\protect\eta \right)}{a_{0}}=10^{22}$ (when $%
\Omega_{g}(\protect\eta)$ is near unity), with $\frac{a_{0}}{a_{3}}=11$,
and $\frac{a_{4}}{a_{0}}=10^{6}$ (to fulfill condition (\ref{cond})).
The right panel shows the power spectrum at the instant 
$\protect\eta $ for which $\frac{a\left(\protect\eta \right) 
}{a_{0}}=10^{16}$ (again when $\Omega_{g}(\protect\eta)$ is near unity), 
with $\frac{a_{0}}{a_{3}}=11$, and $\frac{a_{4}}{a_{0}}=10^{6}$. For the 
sake of comparison the power spectrum of the three-stage scenario 
(De Sitter inflation, radiation, dust with $l = 2$) is also shown 
in both panels (dot-dashed line). Notice the difference in 
slopes at the dust era.}
\label{spec}
\end{figure}

\subsubsection{Topological defects}

Up to now we assumed that at the end of the dark energy era the Universe
will evolve as if it became dominated by dust once again. Nevertheless if
the expansion achieved in the accelerated phase were large enough, either
cosmic strings, or domain walls, or a cosmological constant will take over
instead. We will not consider, however, cosmic strings (whose equation of
state is $p_{K}=-\frac{1}{3}\rho _{K}$) for, as pointed out by Maia \cite%
{Maia93}, it seems problematic to define an adiabatic vacuum in an era
dominated by these topological defects since the creation--anihilation
operators, $A_{(k)}$, $A_{(k)}^{\dag }$, fail to satisfy the conmutation
relations (i. e., condition (\ref{condcuan})) in the range of frequencies
where one should expect RGWs amplification. In short, our approach, as it
is, does not apply to this case.

As for domain walls (topological stable defects of second order with
equation of state $p_{dw}=-\frac{2}{3}\rho _{dw}$ and energy density that
varies as $a^{-1}(\eta)$ -see e.g., \cite{Kolb,shellard}), once the dark
energy evolved as pressureless matter at $\eta =\eta _{4}$ the scale factor
may be approximated by 
\[
a(\eta >\eta _{4})=4a_{4}\left( a_{4}H_{4}\right) ^{-2}\left( -\eta +\eta
_{4}+\frac{2}{a_{4}H_{4}}\right) ^{-2}, 
\]%
so long as $a^{-1}\gg $ $a^{-3}$. That is to say, for $\eta >\eta _{4}$ the
expansion of the Universe is again accelerated whereby $a(\eta )H(\eta )$
resumes growing. The RGWs will be leaving the Hubble radius as soon as $c
H^{-1}$ becomes smaller than their wave length, and eventually none of them
will contribute to the spectrum.

Finally, we consider the existence of a positive cosmological constant $%
\Lambda $ with $\rho _{\Lambda }=\Lambda /\left( 8\pi \right) $ and $%
p_{\Lambda }=-\rho _{\Lambda }$. After the dark energy dynamically mimicked
dust the Universe will become dominated by a very tiny cosmological
constant. The corresponding scale factor is 
\[
a(\eta >\eta _{4})=H_{4}^{-1}\left( -\eta +\eta _{4}+\frac{1}{a_{4}H_{4}}%
\right) ^{-1}. 
\]%
Once more, the expansion is accelerated and RGWs will leave the Hubble radius
and eventually none of them will contribute to the spectrum.

\section{Energy density of the gravitational waves}

By integrating the power spectrum $P(\omega )$, defined by Eq. (\ref%
{densquan}), one can obtain the energy density of the RGWs in terms of the
conformal time. Its current value, evaluated from Eq. (\ref{espectr}), can
be aproximated by \cite{Allen}%
\begin{equation}
\rho _{g}(\eta _{0})\simeq \frac{\hbar }{32\pi ^{2}c^{3}}\left( \frac{a_{0}}{%
a_{2}}\right) ^{2}\left( \frac{a_{1}}{a_{0}}\right) ^{8}H_{1}^{6}(2\pi
H_{0})^{-2}.  \label{dens0}
\end{equation}%
\ To study the evolution of $\rho _{g}(\eta )$ from this point onward we
first consider the case $a_{4}H_{4}>a_{2}H_{2}$. In this case, $\rho
_{g}(\eta )$ evolves as (\ref{dens0}) with $H(\eta )$ and $a(\eta )$
substituted by $H_{0}$ and $a_{0}$, respectively, till some instant $\eta
_{f0} $ in the range $\eta _{0}<\eta _{f0}<\eta _{4}$. When $\eta \geqslant
\eta _{f0}$ the RGWs with $\omega (\eta _{f0})<2\pi \left( a_{2}/a(\eta
_{f0})\right) H_{2}$ must no longer be considered in evaluating $\rho _{g}$
as their wave length exceed the Hubble radius.
Consequently 
\begin{equation}
\rho _{g}(\eta _{f0}<\eta <\eta _{4})\simeq \frac{\hbar }{4\pi ^{2}c^{3}}%
\left( \frac{a_{1}}{a(\eta )}\right) ^{4}H_{1}^{4}\ln \left( \frac{a_{1}H_{1}%
}{a(\eta )H(\eta )}\right) .  \label{dens4}
\end{equation}%
For $\eta =\eta _{4}$, the gravitational waves created at the transition dark energy
era-second dust era begin contriubuting to $\rho _{g}$ thereby, {\small 
\begin{eqnarray}
\rho _{g}(\eta _{4} &<&\eta <\eta _{f1})\simeq \rho _{g}(\eta _{4})\left( 
\frac{a_{4}}{a\left( \eta \right) }\right) ^{4}+\frac{9\hbar }{\pi ^{2}c^{3}}%
l^{2l+2}2^{2l-6}\left( \frac{(-2l)!}{(-l)!}\right) ^{2}\left( \frac{a_{1}}{%
a_{0}}\right) ^{8-4l}  \label{densf1} \\
&& \times \left( \frac{a_{2}}{a_{0}}\right) ^{l-1}\left( \frac{a_{3}}{a_{0}}%
\right) ^{l-3+2/l}\left( \frac{a_{4}}{a_{0}}\right) ^{2l-4-4/l}\left( \frac{%
a(\eta )}{a_{4}}\right) ^{2l-6}H_{1}^{6-2l}\frac{(2\pi H(\eta ))^{2l-2}}{%
-2l+2},  \nonumber
\end{eqnarray}%
}where $\rho _{g}(\eta _{4})$ corresponds to Eq.(\ref{dens4}) evaluated at $%
\eta =\eta _{4}$. For $\eta =\eta _{f1}>\eta _{4}$ where $\eta _{f1}$ is
defined by the condition $a(\eta _{f1})H(\eta _{f1})=a_{2}H_{2}$ (see Figure %
\ref{aH}), the garvitational waves which left the Hubble radius at 
$\eta _{f0}$ reenter it, therefore{\small 
\begin{eqnarray}
\rho _{g}(\eta _{f1} &<&\eta <\eta _{f2})\simeq \rho _{g}(\eta _{f1})\left( 
\frac{a_{f1}}{a\left( \eta \right) }\right) ^{4}+\frac{9\hbar }{\pi ^{2}c^{3}%
}l^{2l}2^{2l-8}\left( \frac{(-2l)!}{(-l)!}\right) ^{2}\left( \frac{a_{1}}{%
a_{0}}\right) ^{16-4l}  \label{densf2} \\
&& \times \left( \frac{a_{2}}{a_{0}}\right) ^{l-4}\left( \frac{a_{3}}{a_{0}}%
\right) ^{l-4+4/l}\left( \frac{a_{4}}{a_{0}}\right) ^{2l-8-4/l}\left( \frac{%
a(\eta )}{a_{4}}\right) ^{2l-10}H_{1}^{10-2l}\frac{(2\pi H(\eta ))^{2l-6}}{%
-2l+6},  \nonumber
\end{eqnarray}%
}where $\rho _{g}(\eta _{f1})$ corresponds to Eq.(\ref{densf1}) evaluated at 
$\eta =\eta _{f1}$. For $\eta =\eta _{f2}$, with $\eta _{f2}$ defined by $%
a(\eta _{f2})H(\eta _{f2})=a_{3}H_{3}$ (see Figure \ref{aH}), the gravitational
waves created at the transition first dust era-dark energy era have wave lengths
shorter than the Hubble radius for first time, and from that point on the
density of gravitational waves can be aproximated by{\small 
\begin{eqnarray}
\rho _{g}(\eta &>&\eta _{f2})\simeq \rho _{g}(\eta _{f2})\left( \frac{a_{f2}%
}{a\left( \eta \right) }\right) ^{4}+\frac{27\hbar }{\pi ^{2}c^{3}}%
l^{4l}2^{4l-6}\left( \frac{(-2l)!}{(-l)!}\right) ^{4}\left( \frac{a_{1}}{%
a_{0}}\right) ^{16-8l}  \label{densf} \\
&& \times \left( \frac{a_{2}}{a_{0}}\right) ^{2l-4}\left( \frac{a_{3}}{a_{0}}%
\right) ^{2l-4+4/l}\left( \frac{a_{4}}{a_{0}}\right) ^{4l-8-4/l}\left( \frac{%
a(\eta )}{a_{4}}\right) ^{4l-10}H_{1}^{10-4l}\frac{(2\pi H(\eta ))^{4l-6}}{%
-4l+6},  \nonumber
\end{eqnarray}%
} where $\rho _{g}(\eta _{f2})$ corresponds to Eq.(\ref{densf2}) evaluated
at $ \eta =\eta _{f2}$.

In the case that $a_{4}H_{4}>a_{2}H_{2}$, Eq. (\ref{dens0}) dictates the
evolution of $\rho _{g}$ between $\eta _{0}$ till $\eta _{4}$. Then, from $%
\eta _{4}$ till $\eta _{f2}$, $\rho _{g}$ obeys Eq. (\ref{densf2}) (note
that $\eta _{f1}$ cannot be defined in this case) and from $\eta _{f2}$
onwards $\rho _{g}$ obeys Eq. (\ref{densf}).

A natural restriction on $\rho _{g}(\eta )$ is that it must be lower than
the total energy density of the flat FRW Universe 
\[
\rho (\eta )=\frac{c}{\hbar }\frac{3m_{Pl}^{2}}{8\pi }H^{2}(\eta ), 
\]%
where $m_{Pl}$ stands for the Planck mass.

It seems reasonable to consider $H_{1}\simeq 10^{38}s^{-1}$ as it
corresponds to the GUT model of inflation \cite{bruce,alan}. The redshift $%
\frac{a_{0}}{a_{2}}$, relating the present value of the scale factor with
the scale factor at the transition radiation era-first dust era, may be
taken as $10^{4}$ \cite{Peebles}, and the current value of the Hubble factor 
$H_{0}$ is estimated to be $2.24\times 10^{-18}s^{-1}$ \cite{Spergel}. It
then follows%
\begin{eqnarray*}
\frac{a_{1}}{a_{0}} &=&1.50\times 10^{-29}\left( \frac{a_{0}}{a_{3}}\right)
^{-\frac{1}{4}+\frac{1}{2l}},\text{\qquad }H_{2}=2.24\times 10^{-12}\left( 
\frac{a_{0}}{a_{3}}\right) ^{-\frac{1}{2}+\frac{1}{l}}s^{-1}, \\
H_{3} &=&H_{0}\left( \frac{a_{0}}{a_{3}}\right) ^{1+\frac{1}{l}}s^{-1},\text{%
\qquad }H_{4}=H_{0}\left( \frac{a_{4}}{a_{0}}\right) ^{-1-\frac{1}{l}}s^{-1},
\\
H_{f1} &=&2.24\times 10^{-12}\left( \frac{a_{0}}{a_{3}}\right) ^{-\frac{3}{2}%
+\frac{3}{l}}\left( \frac{a_{4}}{a_{0}}\right) ^{-1+\frac{2}{l}}s^{-1}, \\
H_{f2} &=&2.24\times 10^{-18}\left( \frac{a_{0}}{a_{3}}\right) ^{\frac{3}{l}%
}\left( \frac{a_{4}}{a_{0}}\right) ^{-1+\frac{2}{l}}s^{-1},
\end{eqnarray*}%
where we have used the relation%
\begin{equation}
H(\eta >\eta _{4})=\left( \frac{a_{4}}{a(\eta )}\right) ^{3/2}H_{4},
\label{Hsde}
\end{equation}%
valid in the second dust era (see Eq. (\ref{sclfac})) with $%
a_{f1}H_{f1}=a_{2}H_{2}$ and $a_{f2}H_{f2}=a_{3}H_{3}$ to evaluate $H_{f1}$
(only defined if $a_{4}H_{4}>a_{2}H_{2}$) and $H_{f2}$, respectively. In our
model, there are only three parameters, namely $l\leq -1$, $2.72<\frac{a_{0}%
}{a_{3}}<11$, and $\frac{a_{4}}{a_{0}}$.

We are now in position to evaluate the evolution of the dimensionless density parameter
$ \Omega _{g}(\eta)\equiv \rho _{g}(\eta )/\rho (\eta )$. Its current value
is\ \cite{Allen}%
\[
\Omega _{g0}\simeq \frac{1}{48\pi ^{3}}\frac{G\hbar }{c^{5}}\left( \frac{%
a_{0}}{a_{2}}\right) ^{2}\left( \frac{a_{1}}{a_{0}}\right)
^{8}H_{1}^{6}H_{0}^{-4}, 
\]%
which in our description boils down to 
\begin{equation}
\Omega _{g0}\simeq 2.00\times 10^{-13}\left( \frac{a_{0}}{a_{3}}\right) ^{-2+%
\frac{4}{l}}.
\label{Og0} 
\end{equation}
$\Omega _{g0}$ is much lower than unity for any choice of $\frac{a_{0}}{a_{3}%
}$ and $l$ in the above intervals. At later times $\Omega _{g}$ evolves as 
\begin{equation}
\Omega _{g}(\eta >\eta _{0})=\Omega _{g0}\left( \frac{a(\eta )}{a_{0}}%
\right) ^{-2+\frac{4}{l}},  \label{omf0}
\end{equation}%
where we have used $H(\eta _{4}>\eta >\eta _{0})=\left( \frac{a(\eta )}{a_{0}%
}\right) ^{-1-\frac{1}{l}}H_{0}$. It is obvious that $\Omega _{g}(\eta )$ is
a decreasing function of $\eta $. For $\eta >\eta _{0}$ we shall distinguish
the two cases mentioned in the previous section.

When condition (\ref{cond}) is fulfilled, the evolution of $\Omega _{g}$ is
given by Eq. (\ref{omf0}) till $\eta =\eta _{f0}$. Then, $\rho _{g}$ changes
in shape from $\eta =\eta _{f0}$ till $\eta =\eta _{4}$, as we have seen.
Consequently, 
\begin{eqnarray}
\Omega _{g}(\eta _{4} &>&\eta >\eta _{f0})\simeq \frac{2}{3\pi }\frac{G\hbar 
}{c^{5}}\left( \frac{a_{0}}{a_{3}}\right) ^{-1+\frac{2}{l}}\frac{a_{2}}{a_{0}%
}\left( \frac{a(\eta )}{a_{0}}\right) ^{-2+\frac{2}{l}}H_{1}^{2}  \label{om4}
\\
&& \times \ln \left[ \left( \frac{a_{2}}{a_{0}}\right) ^{\frac{1}{4}}\left( 
\frac{a_{0}}{a_{3}}\right) ^{\frac{1}{4}-\frac{1}{2l}}\left( \frac{a(\eta )}{%
a_{0}}\right) ^{\frac{1}{l}}H_{1}^{\frac{3}{2}}H_{0}^{-\frac{1}{2}}\right] ,
\nonumber
\end{eqnarray}%
is a decreasing function of $\eta $. Finally, from $\eta _{4}$ on, $\Omega
_{g}$ evolves as 
\begin{eqnarray}
\Omega _{g}(\eta _{f1} &>&\eta >\eta _{4})\simeq \Omega _{g}(\eta
_{4})\left( \frac{a_{4}}{a(\eta )}\right) +\frac{24}{\pi }\frac{G\hbar }{%
c^{5}}l^{2l+2}2^{2l-6}\left( \frac{(-2l)!}{(-l)!}\right) ^{2}\left( \frac{%
a_{1}}{a_{0}}\right) ^{8-4l}  \label{omf1} \\
&& \times \left( \frac{a_{2}}{a_{0}}\right) ^{l-1}\left( \frac{a_{3}}{a_{0}}%
\right) ^{l-3+2/l}\left( \frac{a_{4}}{a_{0}}\right) ^{2l-4-4/l}\left( \frac{%
a(\eta )}{a_{4}}\right) ^{2l-6}H_{1}^{6-2l}\frac{(2\pi )^{2l-2}}{-2l+2}%
\left( H(\eta )\right) ^{2l-4},  \nonumber
\end{eqnarray}

\begin{eqnarray}
\Omega _{g}(\eta _{f2} &>&\eta >\eta _{f1})\simeq \Omega _{g}(\eta
_{f1})\left( \frac{a_{f1}}{a(\eta )}\right) +\frac{24}{\pi }\frac{G\hbar }{%
c^{5}}l^{2l}2^{2l-8}\left( \frac{(-2l)!}{(-l)!}\right) ^{2}\left( \frac{a_{1}%
}{a_{0}}\right) ^{16-4l}  \label{omf2} \\
&& \times \left( \frac{a_{2}}{a_{0}}\right) ^{l-4}\left( \frac{a_{3}}{a_{0}}%
\right) ^{l-4+4/l}\left( \frac{a_{4}}{a_{0}}\right) ^{2l-8-4/l}\left( \frac{%
a(\eta )}{a_{4}}\right) ^{2l-10}H_{1}^{10-2l}\frac{(2\pi )^{2l-6}}{-2l+6}%
\left( H(\eta )\right) ^{2l-8},  \nonumber
\end{eqnarray}

\begin{eqnarray}
\Omega _{g}(\eta &>&\eta _{f2})\simeq \Omega _{g}(\eta _{f2})\left( \frac{%
a_{f2}}{a(\eta )}\right) +\frac{72}{\pi }\frac{G\hbar }{c^{5}}%
l^{4l}2^{4l-6}\left( \frac{(-2l)!}{(-l)!}\right) ^{4}\left( \frac{a_{1}}{%
a_{0}}\right) ^{16-8l}  \label{omf} \\
&& \times \left( \frac{a_{2}}{a_{0}}\right) ^{2l-4}\left( \frac{a_{3}}{a_{0}}%
\right) ^{2l-4+4/l}\left( \frac{a_{4}}{a_{0}}\right) ^{4l-8-4/l}\left( \frac{%
a(\eta )}{a_{4}}\right) ^{4l-10}H_{1}^{10-4l}\frac{(2\pi )^{4l-6}}{-4l+6}%
\left( H(\eta )\right) ^{4l-8}.  \nonumber
\end{eqnarray}%
As follows from (\ref{Hsde}) and (\ref{omf1})-(\ref{omf}), in each case, the
first term redshifts with expansion while the second term of $\Omega _{g}$
grows with $\eta $ (recall that $l \leq -1$). Conditions $\Omega_{g}(\eta
_{f1}>\eta >\eta _{4})\ll 1$ and $\Omega _{g}(\eta _{f2}>\eta >\eta
_{f1})\ll 1$ become just $\Omega _{g}(\eta =\eta _{f1})\ll 1$ and $\Omega
_{g}(\eta =\eta _{f2})\ll 1$, which are true in either case whatever the
choice of parameters. Finally, from Eq. (\ref{omf}) we may conclude that
at some future time larger than $\eta _{f2}$ the condition $\Omega _{g}\ll 1$
will no longer be fulfilled and the linear aproximation in which our
approach rests will cease to apply.

In the opposite case, when condition (\ref{cond}) is not fulfilled, $%
\Omega_{g}$ also grows following Eq. (\ref{omf0}) till $\eta =\eta _{4}$, in
the interval $\eta _{4}<\eta <\eta _{f2}$ it grows according to Eq. (\ref%
{omf2}), and according to Eq. (\ref{omf}) from $\eta _{f2}$ onwards. Our
conclusions of the first case regarding the evolution of $\Omega _{g}$
during these time intervals still hold true.

\begin{figure}[tbp]
\includegraphics*[scale=0.6,angle=-90]{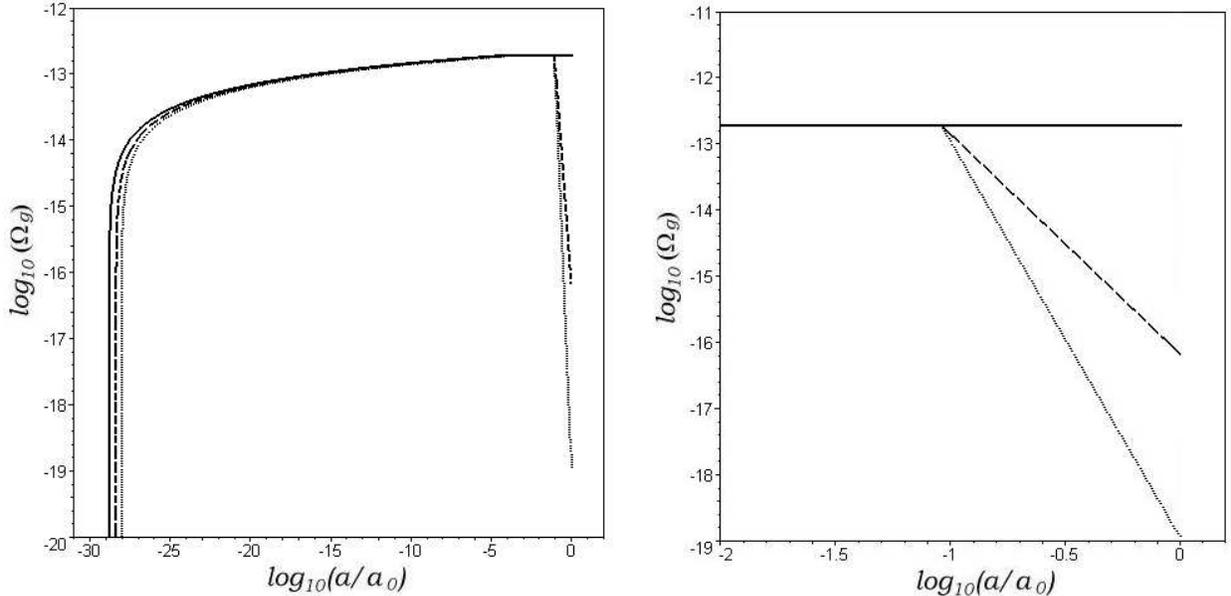}
\caption{Left panel: Evolution of the density parameter $\Omega_g$ with the scale 
factor from the beginning of the radiation era till the present time. The solid line
shows the density parameter predicted by the three-stage model (De Sitter 
inflation, radiation, dust). The dashed and dotted lines depict the density
parameter predicted by a four-stage model having a dark energy era right
after the dust era with $l=-3,\, -1$, respectively, and $a_{0}/a_{3}=11$. Note
that the four-stage model predicts a lower present value for $\Omega_g$ than
the three-stage one as $\Omega_g$ decreases, in a $l$ dependent fashion, 
during the dark energy era. The right panel is a blow up of the region in which
the four stage models ($l=-3,\, -1$),  notably differ from the three stage one.}
\label{Omegag1}
\end{figure}

The behavior of the density parameter $\Omega_{g}$ differs from one
scenario to the other. In the three stage model (inflation, radiation, dust) 
$\Omega_{g}$ remains constant during the dust era \cite{Allen}. In a four
stage model, with a dark energy era right after the conventional dust era, $%
\Omega_{g}$ sharply decreases in a $l$ dependent way during the fourth 
(accelerated) era since long wavelenghts are continuously exiting the 
Hubble sphere \cite{Chiba}, see Fig. \ref{Omegag1}.

\begin{figure}[tbp]
\includegraphics*[scale=0.7]{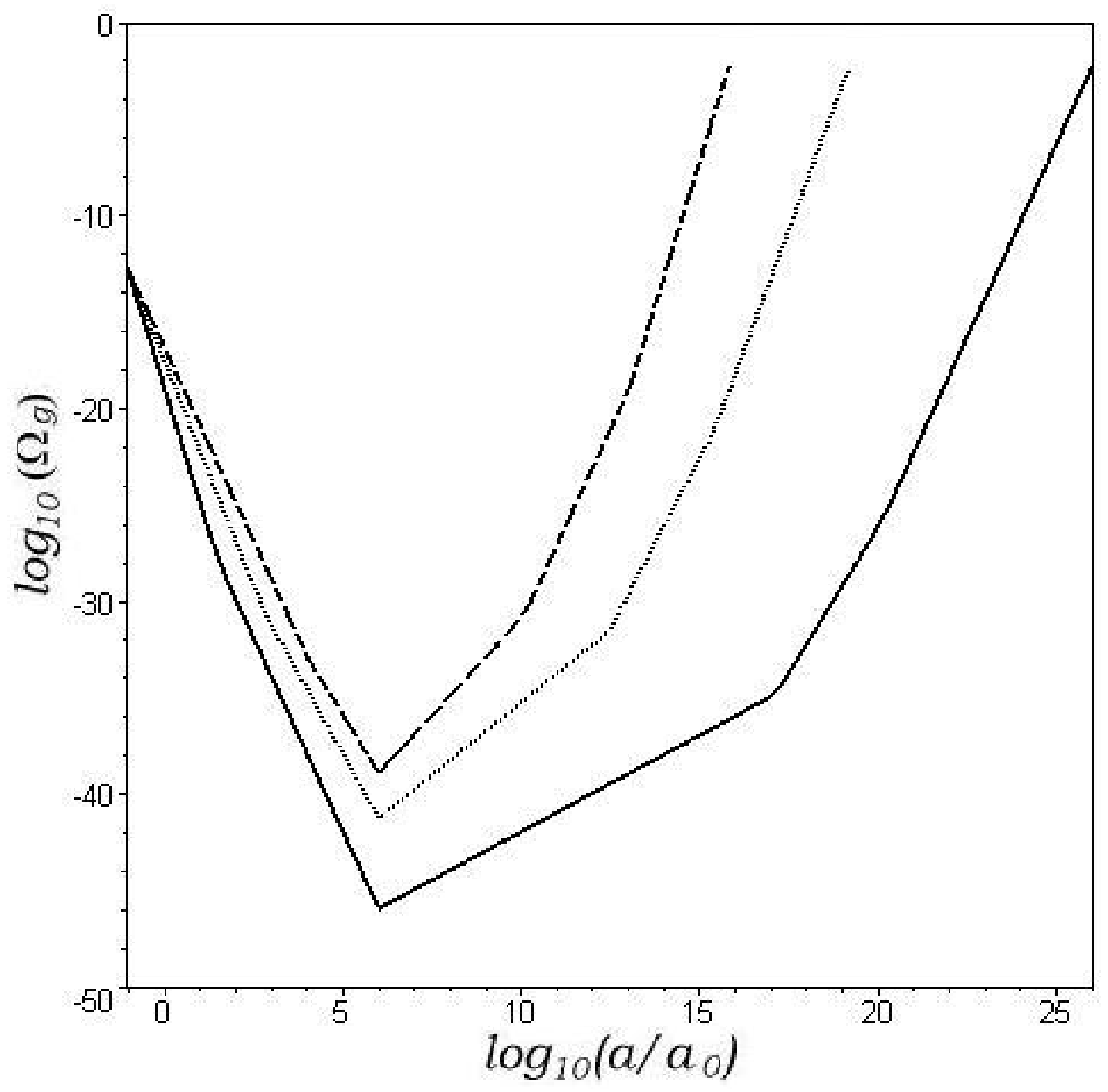}
\caption{Evolution of the density parameter $\Omega_g$ given by Eqs. (\ref{Og0})-(\ref{omf})
with the scale factor in a five-stage scenario 
(De Sitter inflation-radiation-dust-dark energy-second dust era) from the beginning
of the dark energy era. The second dust era starts at the time in which the graphs 
attain their minimum. The solid, dotted and dashed lines correspond to $l=-1$, $l=-1.5$ 
and $l=-2$, respectively. The density parameter decreases in the dark energy era and 
increases in the second dust era till its  value is comparable to unity. From 
that point on our approach (being linear) ceases to apply. We have chosen 
a dark energy era which begins when $a_{3}/a_{0}=1/11$ and ends when 
$a_{4}/a_{0}=10^{6}$.}
\label{Omegag2}
\end{figure}

By contrast, if a second dust era followed the accelerated (dark energy) era, 
$\Omega_{g}$ would grow in this second dust era because long wavelenghts 
would continuously be entering that sphere, see Fig. \ref{Omegag2}. This
immediately suggests a criterion to be used by future observers to ascertain
whether the era they are living in is still our accelerated, dark
energy-dominated, era or a subsequent non-accelerated era. By measuring $%
\Omega_{g}$ at conveniently spaced instants they shall be able to tell.
Further, if that era were the accelerated one and the $\Omega_{g}$
measurements were accurate enough they will be able to find out 
the value of the parameter $l$ occurring in the expansion law 
given by Eq. (\ref{sclfac}). The lower $l$, the higher the slope of $\Omega_{g}$
in the second dust era.

One may argue, however, that these observers may know more easily from
supernova data. Nonetheless, if this epoch lays in the far away future it
may well happen that by then the ability of galaxies to generate stars (and
hence enough supernovae) has seriously gone down and as a result this prime
method might be unavalaible or severely impared. At any rate, even if there
were plenty of supernova, the simple gravitational wave method just outlined
could still play a complementary role.

\section{Discussion}

We have studied the power spectrum and the energy density evolution of the
relic gravitational waves generated at the big bang by considering the
transitions between sucessive stages of the Universe expansion. In
particular, we have studied the effect of the present phase of accelerated
expansion as well as a hypothetical second dust phase that may come right
after the present one. As it turns out, the power spectrum at the current
accelerated era Eq.(\ref{espectr}) formally coincides with the power
spectrum of the usual three-stage scenario. As a consequence, measurements
of $P(\omega)$ will not directly tell us if the Universe expansion is currently
accelerated (as we know from the high redshift supernove data) or
non-accelerated. However, the density parameter of the gravitational waves
evolves differently during these two phases: it stays constant in the
decelerated one and  goes down in the accelerated era in a $l$ dependent manner.
Therefore, the present value of $\Omega_{g}$ may not only confirm us
the current acceleration but also may help determine the value
the parameter $l$ -see Fig. \ref{Omegag1}- and hence give us 
invaluable information about the nature of dark energy.
 
In the far away future measurements of $P(\omega)$, if sufficciently accurate, 
will be able to tell if the Universe is still under accelerated expansion 
(driven by dark energy) or has entered a hypothetical decelerated 
phase (second dust era) suggested by different authors \cite%
{Alam}. This may also be ascertained by measuring the 
density parameter of the gravitational waves at 
different instants to see whether it 
decreases or increases.

\acknowledgments{G.I. acknowledges support from the ``Programa de Formaci\'{o} 
d'Investigadors de la UAB". This work was partially supported by the Spanish 
Ministry of Science and Technolgy under grant BFM2003-06033.}


\begin{thebibliography}{99}
\bibitem{ks} K.S. Thorne, in \textit{300 Hundred Years of Gravitation},
edited by S.W. Hawking and W. Israel (Cambridge University Press, Cambridge,
1987).

\bibitem{alessandra} A. Buonanno, Lecture given at the Theoretical Advanced
Study Institute in Elementary Particle Physics at the University of Colorado
at Boulder, preprint gr-qc/0303085.

\bibitem{bruce} B. Allen, \textit{\textquotedblleft The stochastic
gravity-wave background: sources and detection\textquotedblright}, in \emph{%
Proceedings of the Les Houches School on Astrophysical Sources of
Gravitational Waves}, edited by J.A. Marck \textit{et al.} (Cambridge
University Press, Cambridge, 1996), preprint gr-qc/9604033.

\bibitem{leonid} L.P. Grishchuk, Sov. Phys. JETP \textbf{40}, 409 (1975); Ann. N.Y.
Acad. Sci. \textbf{302}, 439 (1977); Class. Quantum Grav. \textbf{10}, 2449 (1993).

\bibitem{Allen} B. Allen, Phys. Rev. D \textbf{37}, 2078 (1987).

\bibitem{Maia93} M. R. de Garcia Maia, Phys. Rev. D \textbf{48}, 647 (1993).

\bibitem{german} G. Izquierdo and D. Pav\'{o}n, Phys. Rev. D {8}, 12400 5
(2003).

\bibitem{transparent} Ch. Misner, Nature \textbf{214}, 40 (1967); L.P.
Grischuk and A.G. Polnarev in \textit{General Relativity and Gravitation}
vol. 2, edited by A. Held (Plenum, New York, 1980); V. M\'{e}ndez, D. Pav%
\'{o}n and J.M. Salim, Class. Quantum Grav. \textbf{14}, 77 (1997).

\bibitem{lisa} http://lisa.jpl.nasa.gov

\bibitem{snia} A. Riess \textit{et al.}, Astron. J. \textbf{116}, 1009
(1998); P. Garnavich \textit{et al}., Astrophys. J. \textbf{509}, 74 (1998);
S. Perlmutter \textit{et al.}, Astrophys. J. \textbf{517}, 565 (1999); D.N.
Spergel \textit{et al.}, Astrophys. J. Suppl. \textbf{148}, 175 (2003); R.A.
Knop \textit{et al.}, preprint astro-ph/0309368.

\bibitem{iap} \emph{Proceedings of the I.A.P. Conference ``On the Nature of
Dark Energy"}, Paris 2002, edited by P. Brax, J. Martin and J.P. Uzan
(Frontier Group, Paris 2002);\newline
S. Carroll, preprint astro-ph/0310342; V. Sahni, Class. Quantum Grav. \textbf{19}, 3435 (2002); 
V. Sahni, in Proceedings of the Second Aegean Summer School on the Early Universe, Syros,
Greece, 2003, preprint astro-ph/0403324.

\bibitem{Alam} U. Alam, V. Sahni, A. A. Starobinsky, JCAP \textbf{0304}, 02
(2003); M. Sami, T. Padmanabhan, Phys. Rev. D \textbf{67}, 083509 (2003); R.
Kallosh and A. Linde, Phys. Rev. D \textbf{67}, 023510 (2003); V. Sahni and
Yu. V, Shtanov, JCAP \textbf{0311}, 014 (2003).

\bibitem{Parker} L. H. Ford, L. Parker, Phys. Rev. D \textbf{16}, 1601
(1977).

\bibitem{Birrell} N. D. Birrell, P. C. W. Davies, \textit{Quantum Fields in
Curved Space} (Cambridge University Press, 1982).

\bibitem{Abr} M. Abramowitz and I.A. Stegun, eds., \textit{Handbook of
Mathematical Functions}(Dover, New York, 1972).

\bibitem{luca} L. Amendola, Mon. Not. Roy. Astron. Soc. \textbf{342}, 221
(2003), preprint astro-ph/0209494; L.P. Chimento, A.S. Jakubi, D. Pav\'{o}n
and W. Zimdahl, Phys. Rev. D \textbf{67}, 083513 (2003).

\bibitem{Chiba} H. Tashiro, T. Chiba and M. Sasaki, Class. Quantum Grav. 
\textbf{21}, 1761 (2004).

\bibitem{Peebles} P. J. E. Peebles, \textit{Principles of Physical Cosmology}
(Princeton University Press, Princeton, 1993).

\bibitem{string/M} J.M. Cline, J. High Energy Physics \textbf{8}, 35 (2001).

\bibitem{Kolb} E. W. Kolb, M. S. Turner, \textit{The Early Universe}
(Addison-Wesley, Redwood City, 1990).

\bibitem{shellard} A. Vilenkin and E. Shellard, \textit{Cosmic Strings and
other Topological Defects} (Cambridge University Press, Cambridge, 1994).

\bibitem{alan} A.H. Guth, Phys. Rev. D \textbf{23}, 347 (1981).

\bibitem{Spergel} D. N. Spergel \textit{et al.}, Astrophys. J. Suppl. 
\textbf{148}, 175 (2003).
\end{thebibliography}
\end{document}